\documentclass[11pt,preprint]{aastex}

\shorttitle{$\pi$~Aquarii in a Quasi-normal Phase}
\shortauthors{Bjorkman et al.}

\begin{document}

\title{A Study of $\pi$ Aquarii During a Quasi-normal Star Phase:
Refined Fundamental Parameters and Evidence for Binarity}

\author{Karen S. Bjorkman and Anatoly S. Miroshnichenko}
\affil{Ritter Observatory, Dept.\ of Physics \& Astronomy,
       University of Toledo, \\
       Toledo, OH 43606-3390  USA}
\email{karen@astro.utoledo.edu, anatoly@physics.utoledo.edu}

\author{David McDavid}
\affil{Limber Observatory, PO Box 63599, Pipe Creek, TX 78063-3599 USA}
\email{mcdavid@limber.org}

\and

\author{Tatiana M. Pogrosheva}
\affil{Sternberg Astronomical Institute,
Universitetskij pr.\ 13, Moscow 119899, Russia}

\begin{abstract}
We present the results of recent multicolor photometric and
high-resolution spectroscopic observations of the bright Be
star $\pi$~Aquarii.
Observational data collected from the literature were used to study
the star's variations over the last four decades.  The star is
identified with the IR sources F22227$+$0107 in the $IRAS$ Faint Point
Source catalog and $MSX5\_$G066.0066$-$44.7392 in the $MSX$ catalog.
The variations in near-IR brightness
of $\pi$~Aqr are found to be among the largest reported for Be stars.
Since 1996, the star has shown only weak
signs of circumstellar emission, which has allowed us to refine the
fundamental stellar parameters: A$_{V}$=0.15 mag., T$_{\rm eff}$=24000 K,
$\log g$= 3.9, and M$_{V}$=-2.95 mag.
A weak emission component of the H$\alpha$ line has been detected during
the recent quasi-normal star phase.  From analysis of the H$\alpha$
line profiles, we find anti-phased radial velocity variations of the
emission component and the photospheric absorption, with a period of
84.1 days and semi-amplitudes of 101.4 and 16.7~km~s$^{-1}$,
respectively.  This result suggests that $\pi$~Aqr may be a binary
system consisting of stars with masses of $M_1\,\sin^{3}\,i=12.4~M_{\odot}$,
$M_{2}\,sin^{3}\,i=2.0~M_{\odot}$.  We also estimate
the orbital inclination angle to be between 50 and 75\arcdeg.
We suggest that the photometric, spectroscopic, and polarimetric
variations observed during the second half of the $20^{\rm th}$
century may be due to variable mass transfer between the binary
components.
\end{abstract}


\section{Introduction}
\label{intro}

$\pi$~Aquarii (HR 8539, HD 212571) is a bright rapidly rotating
($v\,\sin\,i\sim300~{\rm km~s^{-1}}$) Be star whose variable
characteristics have been noted since the beginning of the
$20^{\rm th}$ century.  However, its behavior until the 1960's is only
well-documented spectroscopically.  \citet{mcl62} reported variations
of the Balmer line profiles (H$\beta$ and H$\gamma$), which appeared
double-peaked most of the time, between 1911 and 1961.  He reported
strong $V/R$ changes ranging from 0.5 to 4.0, and several periods with
an absence of bright emission lines (1936--1937, 1944--1945, 1950).
This study showed that the star was active, but it did not determine
the physical characteristics of the star and its envelope.

Since the late 1950's the star has been frequently observed with
various techniques (multicolor photometry, high-resolution
spectroscopy, polarimetry) that allow a quantitative study.  Several
attempts have been made to determine the fundamental parameters of the
star.  These have resulted, however, in different values being derived,
mainly due to the influence of the circumstellar envelope.
The following ranges of the parameters were reported in the literature:
T$_{\rm eff}$=22500--30000 K, $\log g$= 3.3--4.0,
M$_{V}=-$3.00 -- $-3.83$ mag., and A$_{V}$=0.25--0.69 mag. \citep{no74, u79,
sn81, g85, t85, k89, fr90, zb91}.

Some of the envelope's parameters are noted in the literature.  A wind
terminal velocity of 1450~km~s$^{-1}$ and a mass loss rate of
$2.61\times10^{-9}~M_{\odot}~{\rm yr^{-1}}$, which is one of the largest
among the Be stars, were estimated from UV resonance line profiles
\citep{sn81}.  An envelope temperature of $T_{\rm e}\sim15,000$~K was
derived from optical spectrophotometric and IR photometric
observations \citep{no74,ghj74}.  Profiles of the He {\sc i} lines at
4026 and 4471\AA\ and the Mg {\sc ii} line at 4481\AA\ were used to
estimate the inclination angle of the rotation axis to the line of
sight ($i=25$--$30\arcdeg$) and the ratio of the angular velocity
to the critical value $\omega/\omega_{\rm crit}=0.75$
\citep{gc86,r89}.  However, \citet{r89} pointed out that such high
values of $v\,\sin\,i$ and $\omega/\omega_{\rm crit}$ would imply
$i\ge50\arcdeg$.

\citet{h96} noted that $\pi$ Aqr might be an interacting binary because
of its broad and complex H$\alpha$ line profile, observed in 1980's.
A hypothesis of a binary origin for Be stars was introduced by \citet{kh75}.
So far $\sim$ 40 Be stars have been found to be binaries \citep[e.g.,][]{p97,o97}
using mainly spectroscopic and speckle interferometric techniques. Only a few
eclipsing Be binaries, which allow photometric detection, are known. Most known
Be binaries have orbital periods between a few days and a few hundred days.
Since distances toward Be stars usually exceed 100 pc, the angular separation
between the binary companions is relatively small ($\le$ 10 mas). Therefore,
speckle interferometry, with its contemporary threshold of $\sim$50 mas, is
able to detect only long-period systems \citep[e.g.,][]{m97}.
Thus spectroscopy, with its capability of measuring radial velocity (RV)
variations of a few km\,s$^{-1}$, remains the best tool for searching for Be
binaries; however, this requires long-term monitoring. The results of our
spectroscopic observations provide a good example of such a program.

Around 1985 the star's brightness, polarization, and emission-line
strengths began to decrease.  They reached a minimum in 1996 and have
stabilized since then.  However, line profile variations are still
detectable.  Only a few Be stars that show strong brightness and
emission-line variations have been observed in detail, and these
resulted in different interpretations of the phenomenon
\citep{d82,hum98}.  The recent transition of $\pi$~Aqr from a Be star
to a quasi-normal star phase, in connection with our continuous
long-term monitoring of the star, gave us a unique opportunity to
refine our knowledge of this remarkable object and of the physics of
the Be phenomenon.

We collected published information concerning the behavior of $\pi$ Aqr
since the beginning of the last active emission phase, summarized our own
data obtained during the decline and minimum phases, and analyzed the complete
data set. This allowed us to arrive at important conclusions about the
properties of the $\pi$ Aqr system. In this paper we present our general view
of the active emission phase development, report the evidence for binarity of
$\pi$ Aqr, and refine the fundamental parameters of the system's primary.
Other results concerning the polarimetric behavior of the system and its
disk modeling will be reported elsewhere.

Below we describe our observations
(\S~\ref{obs}), discuss the behavior of $\pi$~Aqr during the last four decades
(\S~\ref{behavior} and \S~\ref{current}) and the H$\alpha$ line profile
variations (\S~\ref{havar}), and suggest an interpretation of the observed
phenomena (\S~\ref{discuss}).

\section{Observations}
\label{obs}

Our monitoring of $\pi$~Aqr includes the following observations:  1)
$UBV$ photometry (shown in  Fig.~\ref{f1}) at a robotic 25~cm 
reflecting telescope (1987--1999)
operated by the Automatic Photoelectric Telescope Service
in Arizona \citep{bgh86}; 2) Spectroscopy with a resolution
$R\sim26,000$ in the range 5280--6600\AA\ with an \'echelle spectrograph
at the 1 m telescope of the Ritter Observatory of the University
of Toledo (1996--2000); and 3) $UBVRIJHK$ photometry with a
two-channel photometer-polarimeter \citep{b88} at a 1 m telescope
of the Tien-Shan Observatory in Kazakhstan (August--December 1998).
Spectropolarimetric monitoring \citep[1989--2001,][]{b00} and
broadband $UBVRI$ polarimetry \citep[1985--1998,][]{mcd99} were also
carried out.
The photometric and polarimetric data will be presented and discussed
in detail elsewhere. In this paper we use our spectroscopic data
and the brightness level in the minimum state of the object to refine its
fundamental parameters.

The Ritter data were reduced with IRAF{\footnote {IRAF is distributed
by the National Optical Astronomy Observatories, which are operated by
the Association of Universities for Research in Astronomy, Inc., under
contract with the National Science Foundation.}.  Eighty-three
spectra with signal-to-noise ratios of 50 and higher were obtained
between August 1996 and September 2001.
The dates of the observations at Ritter, along with information about
the emission component of the H$\alpha$ line, are presented in Table~\ref{t1}.

Several observations of the H$\alpha$ line were obtained in 1990--1991 at
Kitt Peak National Observatory (G.~Peters, private communication) and at the 
University of Colorado's Sommers-Bausch Observatory (M.~Allen, private
communication) in support of the Astro-1 mission, when the star was also
observed spectropolarimetrically in the UV region \citep{b91}.
These data are included in Fig.~\ref{f1} and \ref{f3}. Our study also made 
use of the IUE spectra of $\pi$ Aqr obtained between 1985 and 1995, 
which were retrieved from the IUE final archive \citep{rp99}.

\section{Observed Behavior}
\label{behavior}

\subsection{The Last Active Be-Phase}\label{active}

The last active Be-phase of $\pi$~Aqr, during which line emission was clearly
noticeable, probably began in the early 1950's.  Weak double-peaked
emission in H$\alpha$ and no emission in H$\beta$ were detected by
\citet{bb50} in mid-1949, while \citet{mcl62} reported the absence of
bright emission in 1950.  The early broadband polarimetric data
\citep{ck69,ser70} show that the star already had a large polarization
(nearly 1\%) by the end of the 1950's, which gradually increased until
approximately 1985 \citep{mcd86}.

During the period covered by photoelectric photometry data
(1957--present), $\pi$~Aqr displayed mostly smooth and slow changes of
its brightness, reaching a maximum brightness phase between the
mid-1970's and mid-1980's (see Fig.~\ref{f1}).  However, the lack of
available data in 1975--1985 makes the detailed shape of the optical
brightness maximum uncertain.  The near-IR brightness was reported to
be roughly constant from 1970 to 1985.  Nevertheless, a detailed look
at the $K$-band light curve (Fig.~\ref{f1}b) shows its similarity to
the $V$-band curve.  From all the data displayed in
Fig.~\ref{f1}, we estimate that $\pi$~Aqr reached the peak of its
active Be star phase in approximately 1980--1985, at which time it had
reached a maximum in optical and IR brightness, optical polarization,
and emission strength of the H$\alpha$ line.

The IR excess at this maximum phase was strong and similar to that of
other Be stars, i.e., due to free-free and free-bound radiation from
the circumstellar gas.  Surprisingly, the star was not reported among
the Be stars detected by the $IRAS$ satellite \citep[e.g.,][]{wcl87}.
Since its galactic latitude is rather high ($b=-45\arcdeg$), we
suspected it might have been recorded in the $IRAS$ Faint Source
Catalog.  Indeed, we found an IR source, F22227$+$0107, whose position
coincided with the optical position of $\pi$~Aqr.  We
used the data obtained in the late 1970's to early 1980's to construct
the active Be-phase spectral energy distribution (SED) in the 
range 0.2--60~$\micron$, using the
averaged data from $IUE$ (1978--1985), optical and IR photometry, and the
$IRAS$ fluxes (see Fig.~\ref{f2}b).  The $IRAS$ fluxes obtained in 1983
are in excellent agreement with the IR photometry by \citet{ghj74}
obtained in 1973.

The spectrum of $\pi$ Aqr was not extensively monitored from 1960 to 1995.
Published results showed that the H$\alpha$ line displayed significant
intensity variations on a time scale of months \citep{gm74}.  The
profile shape was single-peaked from 1973--1989
\citep[e.g.,][]{sr78,af82}.  However, in 1989 it changed to a
double-peaked profile, and only this shape was detected until 1995
\citep{h96,vrm98}.  Characteristic H$\alpha$ line profiles obtained by
different studies during the active Be star phase are displayed in
Fig.~\ref{f3}.  The H$\beta$ and H$\gamma$ line profiles were seen
only as double-peaked during the entire Be star phase
\citep{sct92,hkk88}.  According to \citet{mcl62}, the He~{\sc i}
5876\AA\ line varied similarly to the Balmer lines.  Since 1996, when
our spectroscopic monitoring observations began, only weak signs of
emission have been seen in H$\alpha$, while the He~{\sc i} line has
always appeared in absorption (see Fig.~\ref{f4}--\ref{f5}).  Both the
polarization and IR brightness dropped significantly since 1985
\citep{b94a,mcd99}.  While the polarization data shown here still 
contain an interstellar polarization component, analysis of that
interstellar component \citep{bw02} shows that the intrinsic polarization 
itself is declining throughout this later period.  Our data indicate 
that the star's optical
brightness stopped fading in 1995, the line emission reached a minimum
in 1996, and the optical polarization stabilized in 1995
\citep{mcd99,b00}.  Considering all this information, we conclude that
the most recent active Be-phase of $\pi$~Aqr ended in 1995/6 after
lasting for nearly 40 years.

\subsection{The Current Quasi-Normal Star Phase}
\label{current}

How should we refer to the new phase of $\pi$~Aqr's behavior since
1995/6?  Our simultaneous optical and near-IR photometry, obtained in
August 1998, resulted in the detection of the lowest brightness
level ever reported in all filters.  As seen in Fig. \ref{f2}a,
the SED does not show any notable excess flux with respect to normal stars.
The current polarization wavelength dependence is essentially that of a solely
interstellar contribution \citep{mcd99,bw02}.  However,
weak emission is still seen in the H$\alpha$ line.  Thus, we refer to
the current phase as the ``quasi-normal star'' phase.

Only a few spectral lines that are potential indicators of circumstellar
envelope activity fall into the wavelength ranges covered by the 9
non-overlapping orders of our \'echelle spectra. They are Fe~{\sc ii} 5317\AA,
He~{\sc i} 5876\AA, Na~{\sc i}~D$_{1,2}$, Si~{\sc ii} 6347 \&
6371\AA, Fe~{\sc ii} 6383\AA, and H$\alpha$. There are no signs of
the presence of the Fe~{\sc ii} and Si~{\sc ii} lines within 2\%
of the underlying continuum.  The Na~{\sc i} lines have a full-width
at zero intensity (FWZI) of about 110 km~s$^{-1}$ and an
equivalent width (EW) ratio of 1.3, which is consistent with an
interstellar origin. The two remaining lines, He~{\sc i} 5876\AA\
and H$\alpha$, display detectable profile variations. These
variations are better seen in H$\alpha$, since the He~{\sc i} line
is weaker (see Fig.~\ref{f4}d).

As we mentioned in \S~\ref{intro}, the fundamental parameters of the
underlying star as determined by different authors scatter
significantly. One of the main reasons for the scatter is that all
the studies used data obtained during the active Be phase of the
star. The added effect of the circumstellar envelope distorts both
the continuum SED and the spectral line profiles, making
determination of the star's characteristics uncertain and model
dependent. Thus, it is appropriate to make use of the data
obtained during the quasi-normal star phase for re-estimation of
the main parameters of the star.

The color-index variations of $\pi$~Aqr between the Be and
quasi-normal star phases are typical for the positive correlation
between brightness and emission-line strength noted in many
Be stars \citep[e.g.,][]{d82}. At maximum  brightness $U-B$ was bluer,
while $B-V$ was redder than at minimum (Fig.~\ref{f1}).
This effect is due to the circumstellar contribution to the
overall brightness of the system \citep[see][for a discussion]{har00a}.
The values at minimum brightness ($V$=4.85 mag., $U-B=-0.90$ mag.,
$B-V=-$0.20 mag., $K$=5.45 mag.), the presence of an interstellar
polarization component, and the sharp Na~{\sc i}~D$_{1,2}$
lines in the star's spectrum indicate that the interstellar extinction
toward $\pi$~Aqr is not zero.  The averaged color indices in 1998--1999
give $E(B-V)=0.05\pm0.01$~mag, implying a spectral type of B1~{\sc v}
if we assume a mean interstellar extinction law \citep{sm79}. The near-IR
color indices obtained in August 1998 ($J-H=-$0.07 mag., $H-K=-$0.03 mag.)
give essentially the same result. The 1998 brightness in the near IR
suggests the absence of
additional circumstellar emission in this spectral region.
Analysis of the recently released results from the $MSX$ mission
\citep{e99} shows that $\pi$~Aqr was detected (source
$MSX5C\_$G066.0066$-$44.7392) in only one ($A$ band, centered at
8.28~$\micron$) of the 6 photometric bands (total range from 4 to
21~$\micron$), with a flux of 0.385~Jy. This is nearly 10 times lower
than the average IR flux (interpolated to the $MSX$ $A$-band
wavelength) during the active Be phase. From this information we
can derive the star's fundamental parameters by fitting a theoretical
SED to the observed one.

To do this, we used the following: our photometric data in the region
0.36--2.2~$\micron$ obtained in August 1998; the $IUE$ fluxes in a number
of intervals free of spectral lines from 0.12 to 0.31~$\micron$ obtained
on 1995 May 25 (high-resolution spectra LWP 30769 and SWP 54752); and
the $MSX$ 8.28~$\micron$ flux.  These data provided a determination of the
star's effective temperature ($T_{\rm eff}$) and interstellar
reddening ($A_V$).  The H$\alpha$ and He~{\sc i} 5876\AA\ line
profiles obtained in 1996--2000 (see Figs.~\ref{f4}--\ref{f5}) suggest
that the circumstellar contribution to the flux in the continuum was
extremely small.  The SED constructed from our data (Fig.~\ref{f2}a)
was fitted to \citet{k94} model atmospheres with $T_{\rm eff}$ and
$A_V$ as free parameters (note that this procedure is not sensitive to
the stellar gravity).
The best fit values are $T_{\rm eff}=24000\pm1000$ K and
$A_{V}=0.15\pm$0.03 mag. Adopting a distance of $340^{+105}_{-70}$~pc,
obtained by HIPPARCOS \citep{e97}, we derive the following fundamental
parameters of the star:  $\log(L_{\rm bol}/L_{\sun})=4.1\pm0.3$ and
$R_*=6.1\pm2.5~R_{\sun}$.

The $T_{\rm eff}$ is unlikely to
be less than 23,000~K, because otherwise the corresponding
interstellar reddening would be too small for the observed strength of
the Na~{\sc i} lines and current polarization level ($P_V\sim0.45\%$).
It is seen in Fig.~\ref{f2}c that the $IUE$ data, obtained in 1995 when
$\pi$~Aqr already had very weak emission lines, lie just below the
Kurucz model for $T_{\rm eff}=24,000$~K.  During the active Be
phase, the UV flux level was lower than in 1995 because of the
metallic line blanketing effect and a lesser additional circumstellar
emission than that in the optical and near-IR range.
A noticeable dip due to absorption by Fe~{\sc ii} lines in the circumstellar
disk is seen in Fig.~\ref{f2}d.  Thus even if some amount of line blanketing
was still present in the 1995 $IUE$ spectrum, it is unlikely that the level
of the stellar flux in this region is larger than the theoretical one for
$T_{\rm eff}=25,000$~K.

Comparison with theoretical evolutionary tracks \citep{sh93}
provides an estimate of the star's mass, $M_{*}=11\pm1.5~M_{\odot}$
(Fig.~\ref{f6}), and hence the critical rotational
velocity, $v_{\rm crit}=585^{+185}_{-95}~{\rm km~s^{-1}}$, and
$\log\,g=3.9\pm0.1$. Comparison of the observed SED with that of
the \citet{cs77} model for a B1-type star ($T_{\rm eff}=25,000$ K)
rotating at half the critical rotation speed (which corresponds to
$v\sim260~{\rm km~s^{-1}}$ for the parameters adopted by
\citet{cs77} and is close to the $v\,\sin\,i$ of $\pi$~Aqr) shows
excellent agreement at all considered wavelengths.

An independent way to estimate the stellar parameters is to fit the
observed line profiles to theoretical ones.  For this purpose we can
use the averaged He~{\sc i} 5876\AA\ line profile and the H$\alpha$
line wings, which are unaffected by the traveling emission component
which is discussed below.
The detected regular variations of the H$\alpha$ absorption component
were taken into account.  Note that the H$\alpha$ in our case is
mainly indicative of the star's rotational velocity.  We calculated a
grid of theoretical spectra containing these lines using the radiation
transfer code SYNSPEC \citep{hlj95} for $\log~g=4.0$ and $T_{\rm
eff}$ from 20,000 to 30,000~K, and broadened them by rotation.  The
best fits for the H$\alpha$ line were achieved for
$T_{\rm eff}=25,000\pm2,000$~K and $v\,\sin\,i=250\pm10~{\rm km~s^{-1}}$.
The observed He~{\sc i} line profile turned out to be too deep even for
$T_{\rm eff}=20,000$~K and $v\,\sin\,i=300~{\rm km~s^{-1}}$
(Fig.~\ref{f4}d).  Such a discrepancy between the theoretical and
observed profiles of some He~{\sc i} lines, including the one at
5876\AA, has previously been noted by \citet{shl94}.  This might also
imply that the line, which is formed very close to the stellar surface
where the circumstellar envelope is most dense, is still affected by
the circumstellar contribution.  Also, the average observed profile is
rather noisy because of its weakness and both regular and possible
irregular variations.

As mentioned in \S~\ref{intro}, the inclination of $\pi$~Aqr's rotation axis
once was estimated to be 25--30\arcdeg\ based on an analysis of
several He and Mg absorption lines. \citet{r89} noted that the
$v\,\sin\,i$ of the star and this purported inclination angle
were not consistent with each other. The large variations
of the H$\alpha$ line profile, the double-peaked shape of the
H$\beta$ and H$\gamma$ lines, and a high polarization during the
brightness maximum all suggest a configuration much closer to edge-on.
\citet{h96} detected a double-peaked emission profile
of the Fe~{\sc ii} 5317\AA\ line, which suggests an intermediate
inclination angle. Furthermore, the central depressions in the
Balmer line profiles observed in 1989--1995 (see Fig.~\ref{f3})
are not very strong, indicating that the disk is not seen exactly
edge-on. Calculations by \citet{sa94}, performed with
stellar and envelope parameters close to those we find for
$\pi$~Aqr, suggest that the H$\alpha$ profiles which were observed
in 1989--1995 can be reproduced with inclination angles of about
50--60\arcdeg. Calculations by \citet{hum00} show that the
central depression drops below the continuum in H$\alpha$ profiles
for $i\ge80\arcdeg$. Based on this information, we suggest
that $50\arcdeg\le~i\le75\arcdeg$ for $\pi$~Aqr.

%%% END of the new Sect. 3.2 %%%
\subsection{H$\alpha$ Variations During the Quasi-Normal Star Phase}
\label{havar}

During the quasi-normal star phase, the H$\alpha$ line is seen in
absorption most of the time. The photospheric profile is altered by
variable emission, in which we can identify three distinct components.
Two of them are very weak most of the time ($\sim0.02~I_{\rm cont}$ above
the theoretical photospheric profile) and are located at about
$\pm350~{\rm km~s^{-1}}$ (see Fig.~\ref{f4}c).  On only three
occasions, 1996 August 30 and 1998 September 9 and 29, these
components were observed to be significantly stronger (see
Fig.~\ref{f4}a).  The former two profiles are almost
identical, while the third one has weaker emission peaks.  Another
feature of these profiles is that they show stronger emission at
almost all velocities than is seen in the other spectra.  These two
emission components are reminiscent of the double-peaked profile that
the star displayed during the decline phase in 1989--1995.  However,
the peak separation in 1996--2000 is noticeably larger than it was
previously.

The rest of the H$\alpha$ profiles obtained show an additional
emission component of almost the same width ($\sim200~{\rm
km~s^{-1}}$) traveling inside the photospheric absorption within
$\pm101~{\rm km~s^{-1}}$ (see Table~\ref{t1}).  To locate this
component, we normalized each spectrum to the underlying continuum,
removed the profile regions contaminated with telluric water vapor
lines, and subtracted a theoretical photospheric profile calculated
for the star's fundamental parameters as derived above.
The residual profiles obtained in September--November 1999 are shown
in Fig.~\ref{f5}.  The average
parameters of the traveling emission component are as follows:
$v_{\rm FWZI}\sim400~{\rm km~s^{-1}}$, $v_{\rm FWHM}\sim200~{\rm
km~s^{-1}}$, $I\sim0.08~I_{\rm cont}$.  It is difficult to constrain
the shape of the component's line wings because of blending with the
double-peaked structure; however, the core is rather symmetric.

In addition to the RV variations of the traveling emission component,
the entire H$\alpha$ absorption profile changes its RV.
Since the former occupies an insignificant portion of the latter,
we were able to measure their RVs independently. A technique of matching
symmetric parts of the original and mirrored profiles was used for both
components. When measuring RVs of the absorption wings, the region of the
traveling emission component was excluded. Similarly, the photospheric profile
was subtracted to measure RVs of the traveling emission component.
As a result, we found that both variations are periodic, have the same period
(within the uncertainties listed in Table~\ref{t2}), and are anti-phased.
The measured RVs of the emission and absorption component are plotted against
each other in Fig. \ref{f7}b. The relationship between them is well determined
(a correlation coefficient of 0.9), despite some scatter due to the measurement
errors ($\sim$2--3 km\,s$^{-1}$ for most of the datapoints, up to $\sim$5 km\,s$^{-1}$
for those with the lowest signal-to-noise ratios).

The periodic RV variations of the traveling emission component have been
observed during more than 20 cycles. This indicates that the source of the
H$\alpha$ emission rotates around the star in a stable orbit.
The radius of this orbit can be calculated using the derived period and the
semi-amplitude of the RV curve: $r_{\rm orb} = 169 {\rm R}_{\sun}\,{\rm sin}^{-1} i$,
where $i$ is the inclination angle of the orbit. Alternatively, if we assume that
the H$\alpha$ emission source rotates in the star's disk, we can derive
its orbital radius using Kepler's law,
v(R)=v$_{\rm crit}$\,$({\rm r}/{R_{*}})^{-0.5}$.
The result, 133 R$_{\sun}$\,sin$^{2} i$, is not consistent with the above
r$_{\rm orb}$ at any inclination angle. This indicates that the source is located
outside $\pi$ Aqr's disk.
It seems unlikely that the H$\alpha$ emission source is just a cloud of
gas orbiting about the star. A more plausible explanation of this phenomenon
is that there is a secondary companion, surrounded by a gaseous envelope, in a
stable orbit about the primary. Moreover,
the RV amplitude of the absorption component, which we suppose to be mostly
photospheric, is much smaller than that of the emission component.
Such behavior is expected for spectroscopic binaries, a number of
which are Be stars \citep[e.g.,][]{p97}.

If we assume this is evidence that $\pi$~Aqr is a spectroscopic
binary, we can derive appropriate binary parameters for the system.
First we checked for eccentricity effects by fitting the RV curves to
the general equation of orbital motion.
The RV curves for both the absorption and emission components were
found to be virtually symmetric within the measurement errors, which
indicates that the orbital eccentricity is very small or absent. Thus,
the orbit is approximately circular.

The best fit parameter values are shown in Table~\ref{t2}.  
The RV phase curves for both components are presented in Fig.~\ref{f7}a.
These results strongly argue in favor of a binary origin of
the system, in which the absorption-line spectrum is associated with a
more massive B-type primary, while the traveling emission component in
the H$\alpha$ line originates in a region around a less massive
secondary.
The differences between the parameters derived from the RVs of the H$\alpha$
emission and absorption components are most likely due to individual
measurement errors, quoted above.
We should also note that the RVs of the emission component
depend on the adopted theoretical photospheric profile.  The latter
affects the resulting RV, especially at the extremal positions of the
emission component, because of the photospheric profile curvature.
However, this effect only results in an uncertainty of $\le$ 0.05 days
in the orbital period and of $\le$5 km\,s$^{-1}$ in the RV semi-amplitude.
The emission phenomenon, which gave rise to the three
double-peaked profiles mentioned above, does not have any effect on
the periodicity.  It does not even hide the traveling emission
component.  The only profile in which the traveling component was not
seen is that of 1998 September 29.  The component's RV at that time
should have been about zero, and we were unable to measure it because
of the contamination with additional emission.

Using these orbital parameters, we can estimate the orbiting masses
using corresponding equations for Keplerian motion with zero
eccentricity.  Applying the data from Table~\ref{t2}, one
can calculate the mass function $f(M)=0.041\,M_{\odot}$ and,
using the RV semi-amplitudes, $M_1\,sin^{3}\,i=12.4~M_{\odot}$,
$M_{2}\,sin^{3}\,i=2.0 M_{\odot}$, and $M_{2}/M_1=0.16$.

The equivalent width of the traveling emission component listed in
Table~\ref{t2} also shows variations with the 84 day period.
Additionally, the component strength seems to decrease with time.  The
phase curve corrected for the temporal trend shows that the emission
becomes weaker at phases 0.6--0.7 (Fig.~\ref{f7}c).  This phase
interval is centered at the moment when the RV=0, and the secondary
should be in front of the primary.  This phenomenon might be caused by
a non-spherical distribution of circumstellar gas around the
secondary, so that the star occults more matter at the mentioned
phases.  In other words, there is additional circumstellar matter
between the stars.  At the same time, the absence of such an effect
around phase 0.25 suggests that there is no obscuration of the
emitting material by the visible star and its disk remnant.

Thus, our spectra obtained during the current quasi-normal star
phase of $\pi$~Aqr have revealed a cyclic emission activity in the
H$\alpha$ line.  Other indicators of the circumstellar envelope,
except for the very weak double-peaked emission in the H$\alpha$
line, have vanished. RV variations of the photospheric H$\alpha$
profile, indicating an orbital motion of the primary component,
have also been detected. Similar variations are seen in the He
{\sc i} 5876\AA\ line. However, it is more difficult to measure
them since the line is shallower than H$\alpha$.

\section{Discussion}
\label{discuss}

The RV variations of the H$\alpha$ profile components gave another
estimate of the visible star mass ($M_1\,\sin^{3}\,i=12.4~M_{\odot}$).
This estimate is probably only accurate to within 20\% because of the
measurement errors and uncertainties mentioned above.  If we assume
that the orbital inclination angle is the same as the disk inclination
angle ($\sim70\arcdeg$, see \S~\ref{current}), then
$M_1\sim15\pm3~M_{\odot}$.  This is somewhat higher than our
evolutionary track estimate, but still within the calculation
uncertainties.  A more accurate spectroscopic estimate might be
derived given more observations at a higher signal-to-noise ratio and
from a drier site, if the system remains in the quasi-normal star
phase.

The derived $T_{\rm eff}$ is in good agreement with the spectral type
(B1) assigned to the star in earlier studies \citep [e.g.,][]{g68}.
We found that $\pi$~Aqr is basically less luminous than previously
thought, since the star was usually estimated to be brighter because
of the effect of the circumstellar envelope.  This effect also caused
overestimation of the interstellar reddening in most previous papers
(see \S~\ref{intro}).

The measured semi-amplitude ratio of the absorption and emission
component RVs during the quasi-normal star phase suggests that 
the stellar companion, if placed at the center of the H$\alpha$
emission source, should be
$\sim$6 times less massive than the
visible star. According to the derived RV curve parameters and the
estimated orbital inclination angle, the mass associated with the
emission component is between 2 and 3 $M_{\odot}$.  Thus, it is
most likely a star surrounded by a gaseous envelope. Evolutionary
arguments would suggest that this secondary component should be a
main-sequence star. The evolutionary tracks \citep{sh93}
for such a star indicate that it would be an A-type (or even
F-type) star with a luminosity more than 2 orders of
magnitude smaller than that of the primary. This would imply a
visual magnitude difference of at least 4 magnitudes. Therefore,
the contribution from such a secondary to the continuum radiation
of the system is negligible, and it is not surprising that no
signs of it have been detected.

From the estimated brightness ratio of the companions, one can
model the H$\alpha$ profiles observed during the quasi-normal star
phase in order to estimate an average density of the circumstellar
gas around the secondary. For this purpose, we assume that the
emission line due to this gas is described by a Gaussian profile,
and that the secondary's photospheric H$\alpha$ profile is
described by a \citet{k94} model atmosphere with $T_{\rm eff}=9,000$~K
and $\log\,g=4.0$. The secondary's rotation rate and
fundamental parameters are not crucial because of the large
brightness ratio. The results for one observed spectrum are shown
in Fig.~\ref{f8}. The H$\alpha$ emission in the secondary
spectrum should be rather strong (equivalent width of about 20
\AA), which would require a mean number density on the order of
10$^{11}$~cm$^{-3}$, assuming a spherical density distribution.
Such an emission line strength and envelope density is similar to
that of Herbig Ae/Be stars, which have mass loss rates of about
10$^{-8}M_{\odot}$~yr$^{-1}$ \citep{bc95}.
However, this does not mean that the secondary is at the pre-main-sequence
evolutionary stage. Its envelope is most likely formed due to mass exchange
in the system.

The semi-major axis of the binary orbit can be derived from the
data of Table~\ref{t2}, which gives $a=0.96\sin^{-1}i$~A.U.
This value corresponds to an angular separation of $\sim$3~mas at
the HIPPARCOS distance of $\pi$~Aqr. This is far
below the Rayleigh limit of 55 mas reported for the star
by \citet{m97}, who searched for binary components around
Be stars using speckle interferometry.

The observational data for $\pi$~Aqr, collected and summarized in
Fig.~\ref{f1}, show that variations of its brightness,
color-indices, emission-line strength, and polarization correlate
with each other. Three major distinct phases can be recognized in
the stellar behavior during the last four decades:  1) the active
Be star phase, showing an increase of brightness, emission-line
strength, and polarization (early 1950's through $\sim$1985); 2)
the transition phase, with a rapid decrease of these same
characteristics ($\sim$1985 through 1995); and 3) the
quasi-normal star phase, with only very weak signs of
circumstellar activity (since 1996).

The temporal correlation indicates that all the detected variations
can be explained by the same mechanism.  This mechanism is most likely
the added flux due to a circumstellar gaseous envelope with a variable
density, as previously suggested for $\pi$~Aqr by \citet{no77}.
Gradual strengthening of the circumstellar emission until the
mid-1980's suggests that the amount of matter in the disk was
increasing during this time.  Therefore, an SED constructed from data
obtained at a particular time during this phase reflects only the amount
of matter accumulated in the disk up to that time. An SED constructed from
only observations at an earlier or later time would lead to a different
mass estimate for the disk.

The Stokes $QU$ parameters measured at different times are distributed
along a well-established line \citep[see][]{b00}, implying that the
intrinsic polarization position angle (PA) was nearly constant.  According to
current models \citep[e.g.,][]{wbb97}, the intrinsic polarization of Be
stars originates in the circumstellar disk, and the resulting PA is
perpendicular to the disk equatorial plane.  This has been verified
observationally by \citet{q97}.  Thus, the observed intrinsic PA
constancy leads to the conclusion that the orientation of the disk
around $\pi$~Aqr with respect to the line of sight did not change
during the period considered here.

The unusual H$\alpha$ line profile shape observed in 1973--1989 (see
Fig. \ref{f3}) might imply that mass transfer from the primary's
disk into the secondary's Roche lobe through the inner Lagrangian
point $L_1$ occurred during a part of the active Be phase \citep[see
Fig.~6 from][]{h85}.  This suggestion can be supported by the
following arguments:

1) The RVs of the H$\alpha$ profile dominating peak were confined
within $\pm100~{\rm km~s^{-1}}$ (see Fig.~\ref{f3}), which is well
inside the peak separation of the double-peaked structure observed in
1989--1995.  This suggests that the dominating peak was formed in a
low-velocity region of the disk.  If Be star disks are Keplerian
\citep[see][]{hh97}, this region should be close to the disk's outer
parts.

2) The size of the primary's Roche lobe, $\sim0.6\,a$ for the
derived components' mass ratio \citep{pw85}, is $\sim25~R_1$.
This value is similar to the disk size of $\gamma$ Cas derived
from interferometry \citep{b99}. Since the emission-line
spectrum of $\pi$ Aqr during the active Be phase was even stronger
than that of $\gamma$ Cas, the primary's disk in the $\pi$~Aqr system
might fill the primary's Roche lobe.

3) The absence of mass transfer out of the disk would result in a
very different picture of the object's variability since 1989.
In this case, in order to get rid of the bulk of the disk matter,
one needs to accrete it onto the primary. In the beginning, this
process would re-shape the line profiles, making them double-peaked
with a larger intensity, as the amount of matter in the disk is still
constant. However, we observed a decrease of emission peak intensity
only a few months after the H$\alpha$ profile re-shaping occurred.

We should note here that reasons for the mass loss variations from
the primary star are not known. Concerning the disk
dissipation process, if we assume a constant velocity of
1~km~s$^{-1}$, a particle would cross the $\pi$~Aqr
disk in 2 years. This is consistent with the observed timescale of
the disk dissipation ($\sim$5~yr).

A detection of $\pi$~Aqr by the $ROSAT$ survey in 1990 (during the
early decline phase) with $\log(L_{\rm
x}/L_{\rm bol})\sim-6.6$ (corrected using our new value of $L_{\rm
bol}$) shows that its x-ray luminosity is only marginally higher than
the average value for stars hotter than B1-type \citep{bsc96}.
However, the detected x-ray luminosity is only 0.3~dex smaller than
that of $\gamma$~Cas, which has been suspected of having a compact
companion \citep{mws76}, and it may also be variable.  $\pi$~Aqr was
one of only a few Be stars detected by EUVE \citep{c99}, which might
imply the presence of flux in excess of the photospheric contribution
in this wavelength region.  Such an excess might be attributed to
chromospheric activity of the secondary.

A number of other bright Be stars show variations of the H$\alpha$ line
profile, similar to those of $\pi$~Aqr, during their active Be phase
(e.g.\ $\gamma$~Cas, 59~Cyg, $\zeta$~Tau). The binary hypothesis may
well be applied to these objects. $\zeta$~Tau is already known as a
spectroscopic binary with a period of 132.91 days \citep[e.g.,][]{u52},
close to the period we found for $\pi$~Aqr. Recently \citet{har00b}
found a period of 203.59 days in the emission line variations of
$\gamma$~Cas and attributed them to binary orbital motion.

Now we can make some suggestions in response to the questions
raised by \citet{mcd99} about the physical reasons for the
polarization decline in $\pi$ Aqr. The
multiwavelength behavior of $\pi$~Aqr during the quasi-normal star
phase indicates that the primary's disk indeed lost its
replenishment source. The temporal variations of the system indicate
that the disk is not a static structure, but rather has a
variable mass and density distribution. The underlying star does
not seem to change its fundamental characteristics. However, a
triggering mechanism which could turn off the mass loss is not clear
yet. At present we have no information which would allow us to
discuss this subject in any detail.

\section{CONCLUSIONS}

We have presented and analyzed photometric, polarimetric, and
spectroscopic data for the bright  Be star $\pi$~Aqr
obtained during the last four decades. The star showed a
brightening between the late 1950's and early 1970's, a maximum
phase through 1985, a decline between 1985 and 1995, and a
low-brightness phase by the present time. This large data set
allowed us to derive new, more reliable physical parameters of
the star. We also have identified $\pi$~Aqr with IR sources
in the $IRAS$ Faint Source Catalog and in the $MSX$ satellite point
source catalog.

Regular RV variations of a traveling emission component within
the H$\alpha$ line profile and of the underlying photospheric
absorption profile with a period of 84.1 days have been detected
during the quasi-normal star phase of $\pi$~Aqr in 1996--2000. The
combined picture of the profile structure and variations suggests
that $\pi$~Aqr may be a close binary system with variable mass
exchange.

The following estimates of the fundamental parameters of the
system's visual component and the interstellar reddening were
made: $M_1=11\pm1.5~M_{\sun}, R_1=6.1\pm2.5~R_{\sun},
T_{\rm eff}=25,000\pm2,000~K, \log(L_{\rm bol}/L_{\sun})=4.1\pm0.3,
A_V=0.15$~mag. Also, some constraints were placed on
the system inclination angle, binary mass ratio and separation:
$50\arcdeg\le~i\le75\arcdeg, M_{2}/M_1=0.16, a=0.96\,\sin^{-1}\,i$~A.U.
The secondary is most likely an A--F main-sequence star.

There are some major consequences of our interpretation of the
active Be phase.  First, the mass loss rate cannot be determined
accurately from the SED (IR excess) obtained at a 
particular time, because the
star's gradual brightening indicates a disk mass growth with time.
Second, variations of the H$\alpha$
line profiles of other Be stars, which show a structure similar to
that of $\pi$~Aqr during its active Be phase, should be searched
for periodicities.  We would like to emphasize here that we do not
claim that this mechanism is operating for all Be stars; however,
a certain fraction of them may well be explained by a similar
hypothesis.

Further observations of $\pi$~Aqr are of interest in order to
follow the quasi-normal star phase and detect any new possible
brightening. They would help to reconstruct the the disk
replenishment process and improve the parameters of the $\pi$~Aqr
binary system, which was first suspected by \citet{h96} but not
detected with speckle interferometry \citep{m97}.

\acknowledgments

We thank an anonymous referee for comments which helped improve
the manuscript.  We are grateful to R.~W.~Hanuschik for providing 
us with his H$\alpha$ line profiles of $\pi$~Aqr obtained in 
1982--1995, to A.~V.~Mironov for sending us unpublished photometric 
$WBVR$ data obtained in 1986, and to G.~Peters and M.~Allen for 
providing the KPNO and SBO H$\alpha$ spectroscopic observations.  
We thank the present and past members of the Ritter observing team, 
especially N.~Morrison, C.~Mulliss, K.~Gordon, D.~Knauth, and 
W.~Fischer, for assistance in the data acquisition.  We thank 
J.~Bjorkman for useful conversations about these results.  AM 
and KSB acknowledge support from NASA grant NAG5-8054.  Karen 
Bjorkman is a Cottrell Scholar of the Research Corporation, and 
gratefully acknowledges their support.  Support for observational 
research at Ritter Observatory has been provided by The University 
of Toledo, by NSF grant AST-9024802 to B.~W.~Bopp, and by a grant 
from the Fund for Astrophysical Research.  Technical support at 
Ritter is provided by R.~J.~Burmeister.  This research has made use
of NASA's Astrophysics Data System and of the SIMBAD database,
operated at CDS, Strasbourg, France.

%%%%%%%%%%%%%%%%%%%%%%% FIGURE CAPTIONS %%%%%%%%%%%%%%%%%%%%%%%%%%%%%%%%%%%%%%

\clearpage

\begin{figure}
\epsscale{0.8}
\plotone{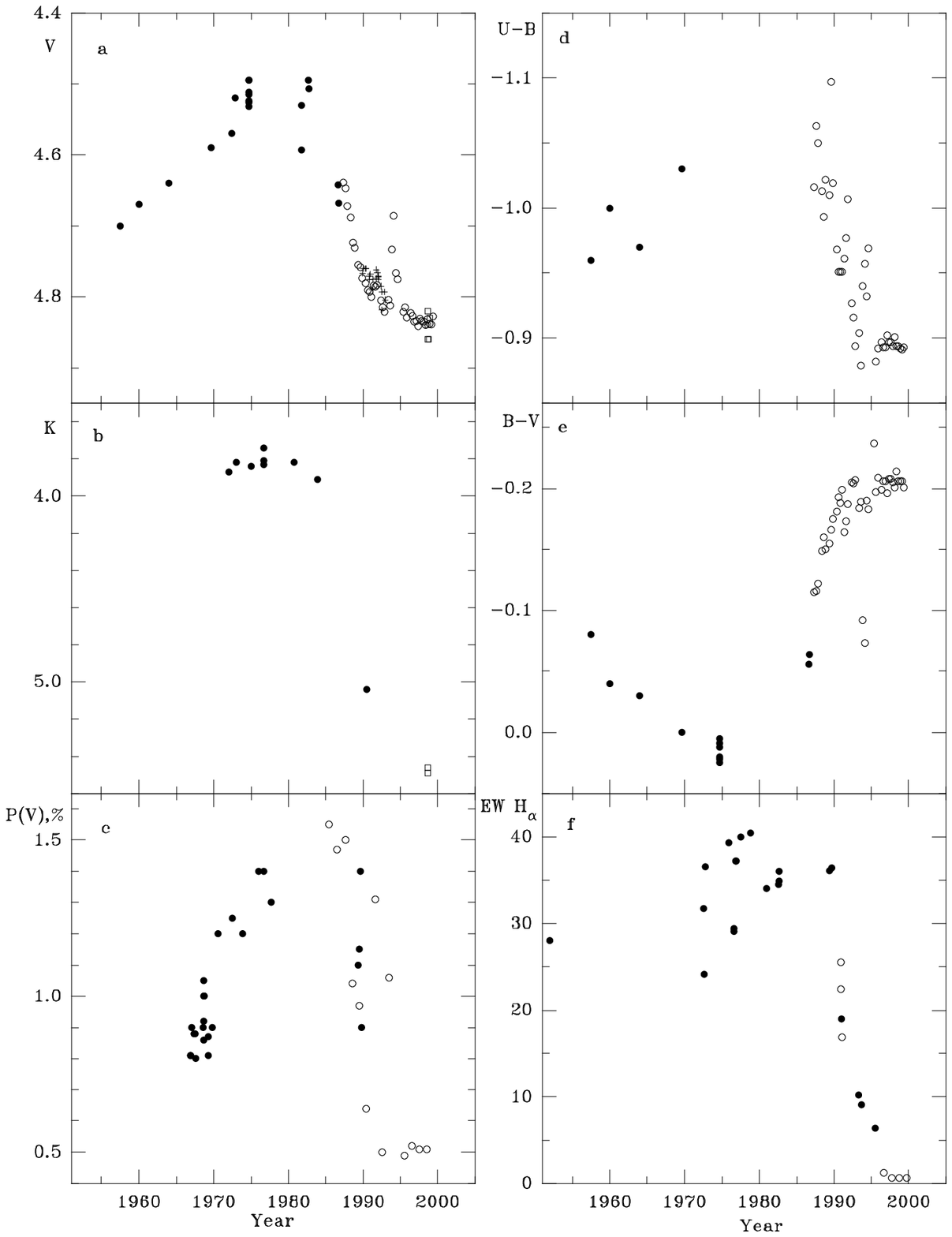}
\caption{Photometric, spectroscopic, and polarimetric
variations of $\pi$~Aqr in 1952--1999.  Data taken from the literature
are marked with filled circles; Arizona photometry (averaged over
every 1/4 year) and Texas polarimetry \citep[from][]{mcd99} with open
circles; Tien-Shan photometry with open squares; and HIPPARCOS data
transformed into $V$ magnitudes, using the approximation by \citet{har98},
with pluses.  Equivalent widths of the H$\alpha$ line are given
in \AA. \label{f1}}
\end{figure}

\clearpage

\begin{figure}
\epsscale{1.0}
\plotone{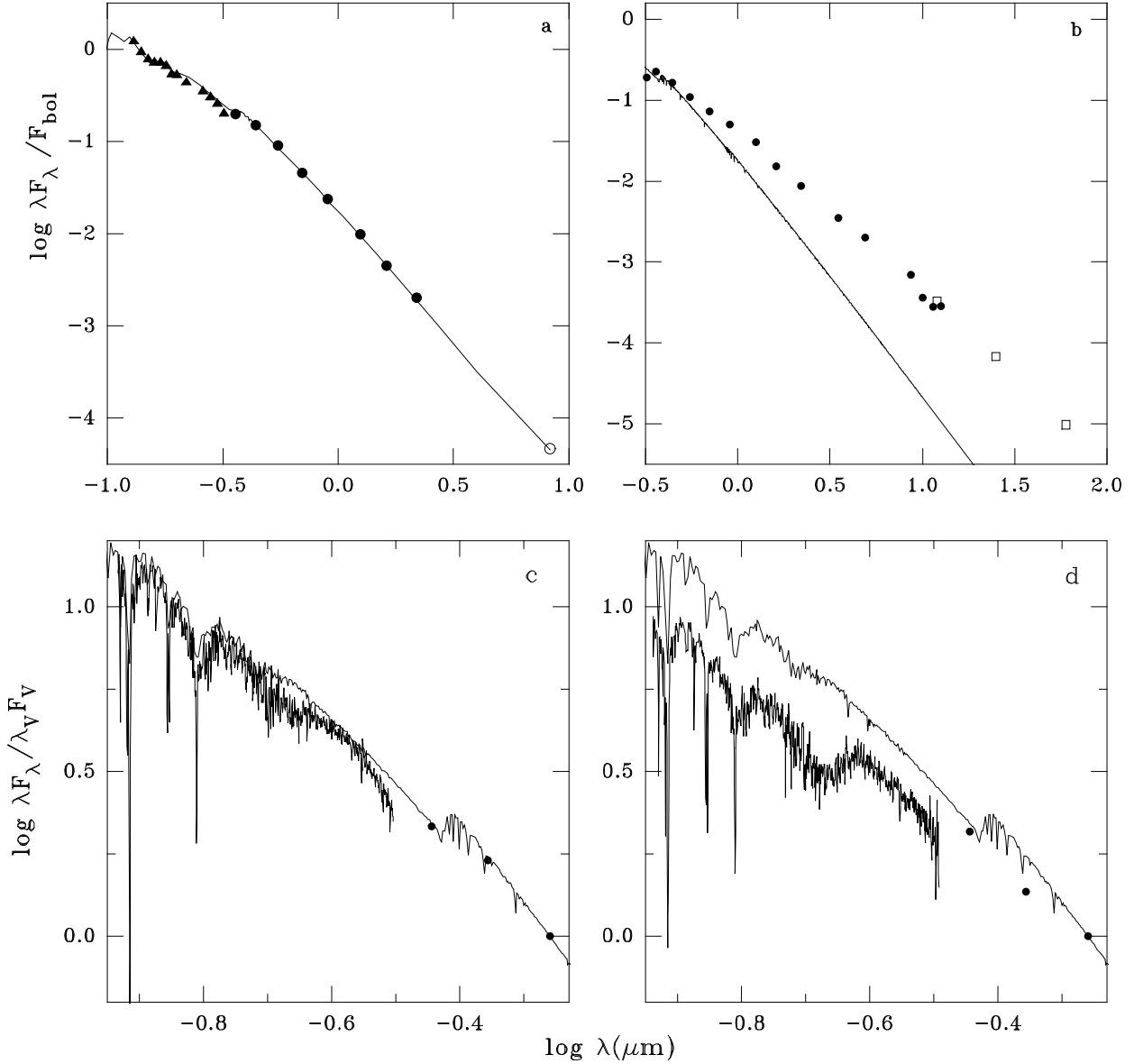}
\caption{Dereddened spectral energy distributions of $\pi$~Aqr in the
active Be (panels b,d) and quasi-normal star (panels a,c) phases.
Ground-based photometric data are shown by filled circles, the $MSX$
(panel a) and $IRAS$ (panel b) data by open circles, and UV continuum
fluxes from the IUE spectrum LWP 30769 by triangles.
The Kurucz model atmosphere for $T_{\rm eff}=25,000$~K and $\log g=4.0$
is shown by solid lines in all panels.
The $IUE$ spectra of $\pi$~Aqr supplemented with the UBV photometric data
in the corresponding phases are shown in panels c and d.  \label{f2}}
\end{figure}

\clearpage

\begin{figure}
\epsscale{0.6}
\plotone{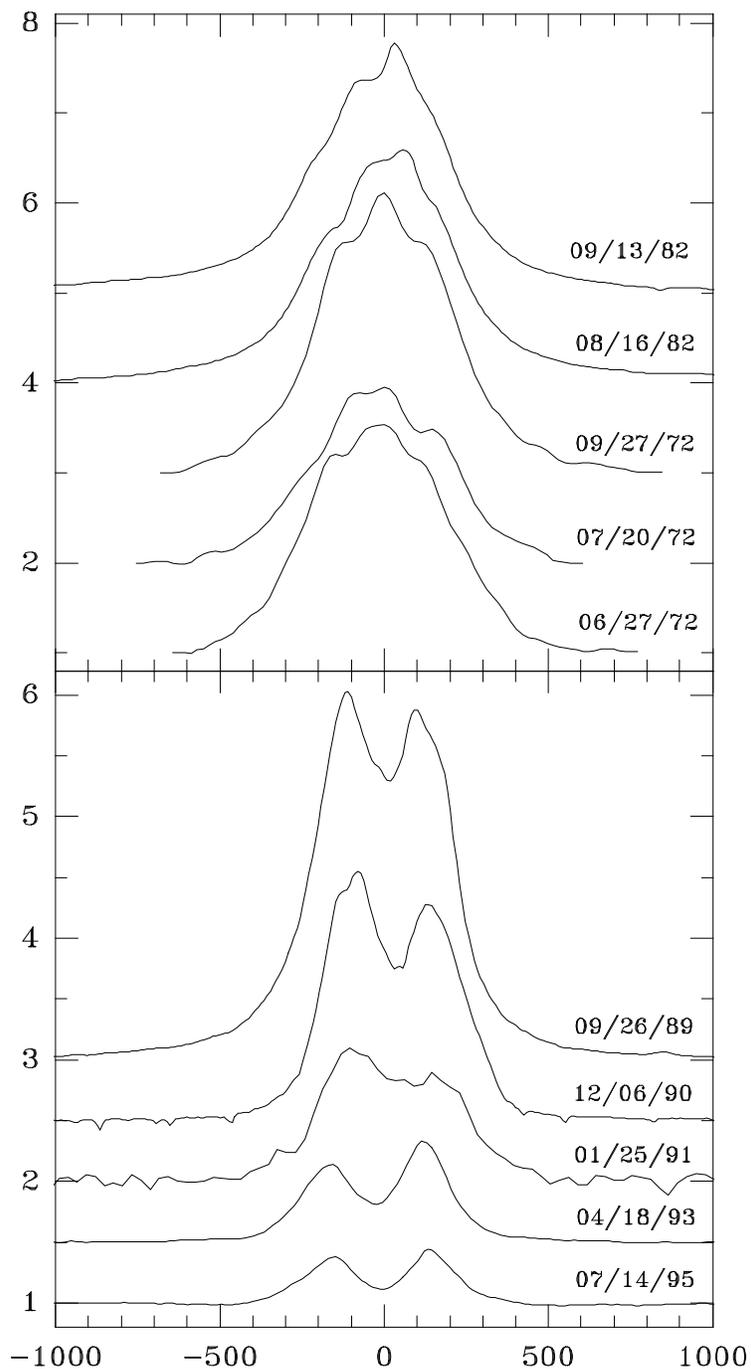}
\caption{Evolution of the H$\alpha$ profile of $\pi$~Aqr
during the last active Be star phase. The profiles obtained in
1972--1989 are shown in the upper panel \citep[from][]{gm74,h96},
while those obtained in 1989--1995 are shown in the lower panel
\citep[from][and this paper]{h96}. The intensities are normalized to
the underlying continuum; the radial velocities are given in km~s$^{-1}$.
\label{f3}}
\end{figure}

\clearpage

\begin{figure}
\epsscale{0.9}
\plotone{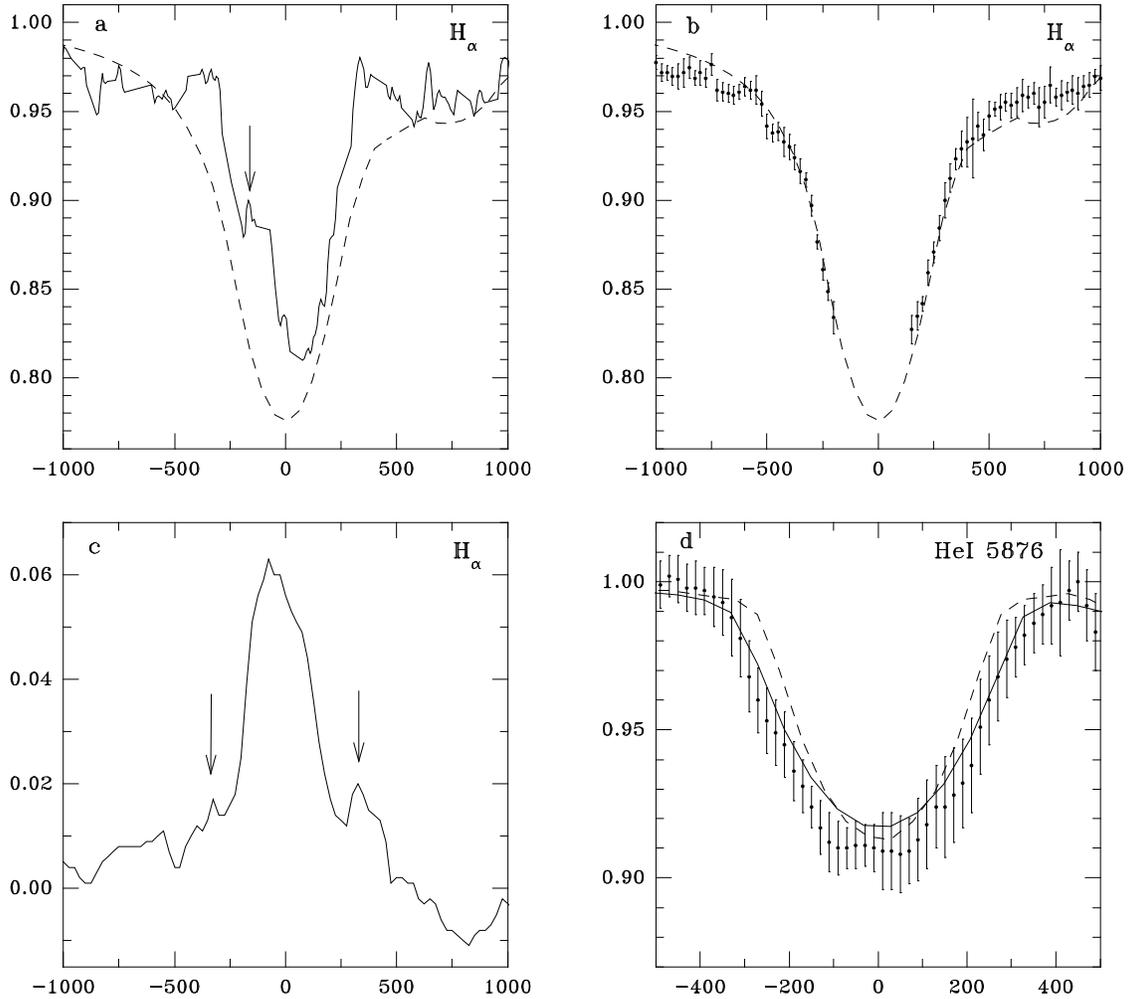}
\caption{Typical line profiles of $\pi$~Aqr during the quasi-normal star
phase. The intensities and radial velocities are in the same units as in
Fig.~\ref{f3}.
a) The averaged H$\alpha$ profile observed during two emission events
in 1996 and 1998. The arrow marks the location of the traveling emission peak.
b) The averaged H$\alpha$ profile with the region affected by emission
extracted. All profiles with the traveling emission peak at velocities
$v\ge75~{\rm km~s^{-1}}$ were used to average the blue wing, while those with
$v\le-95~{\rm km~s^{-1}}$ were used to average the red wing.
The theoretical line profile for $T_{\rm eff}=25,000$~K, $\log~g=4.0$, and
$v\,\sin\,i=250~{\rm km~s^{-1}}$ is shown in both panels by a dashed line.
c) The average profile (with the photospheric contribution subtracted)
calculated from 63 observations obtained in 1996--1999 is shown in the
lower panel.  The arrows mark the double-peaked structure arising in
the visible star's disk.
d) The averaged He {\sc i} 5876\AA\ line profile during the quasi-normal
star phase. The observed averaged profile is shown by filled circles along
with the errors.  The theoretical profile for $T_{\rm eff}=25,000$~K,
$\log~g=4.0$, and $v\,\sin\,i=250~{\rm km~s^{-1}}$ is shown by a
dashed line, while that for $T_{\rm eff}=20,000$~K, $\log~g=4.0$,
and $v\,\sin\,i=300~{\rm km~s^{-1}}$ is shown by a solid line. \label{f4}}
\end{figure}

\clearpage

\begin{figure}
\epsscale{0.3}
\plotone{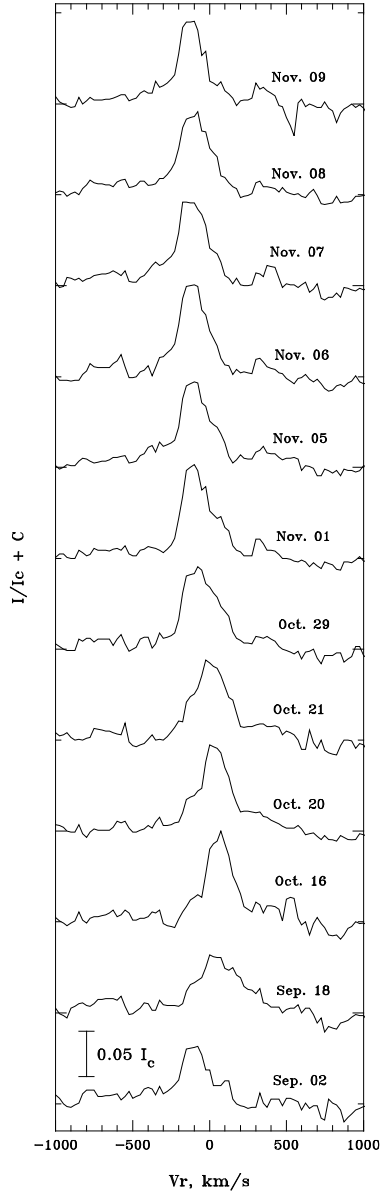}
\caption{Residual H$\alpha$ line profiles obtained in
1999.  The theoretical profile for $T_{\rm eff}=25,000$~K,
$\log~g=4.0$, and $v\,\sin\,i=250~{\rm km~s^{-1}}$ was subtracted from
each observed profile.  Individual profiles are shifted by 0.1~$F_{c}$
with respect to each other.  Intensity and velocity are in the same
units as in Fig.~\ref{f3}.  \label{f5}}
\end{figure}

\clearpage

\begin{figure}
\epsscale{1.0}
\plotone{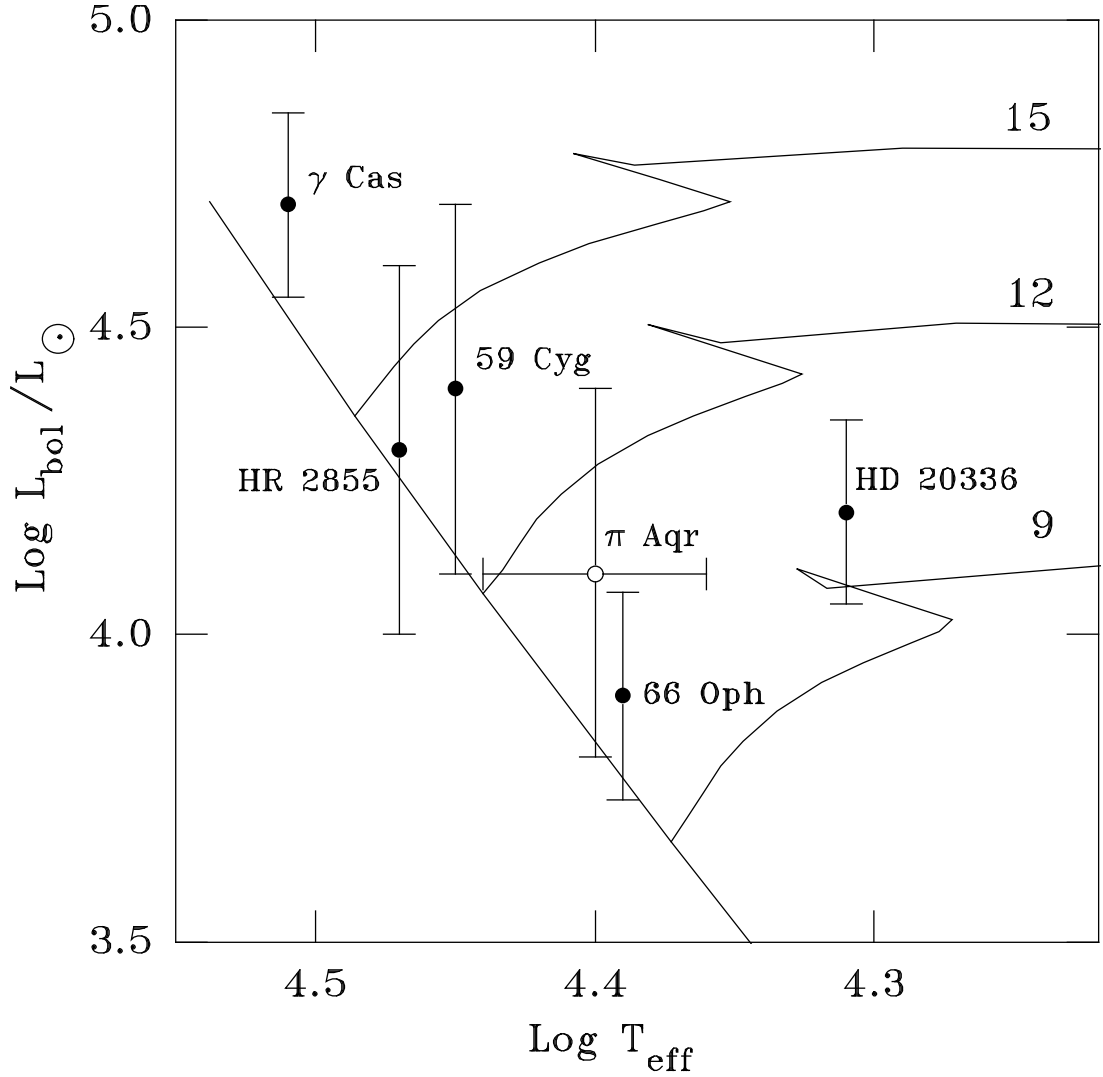}
\caption{Location of $\pi$~Aqr in the Hertzsprung-Russell
diagram.  The open circle with error bars marks our results.
Evolutionary tracks for 9, 12 and 15 $M_{\sun}$ from \citet{sh93}
are shown by solid lines. The positions of several other Be stars
\citep[from][]{zb91} with similar spectral types are given for
comparison. \label{f6}}
\end{figure}

\clearpage

\begin{figure}
\epsscale{1.0}
\plotone{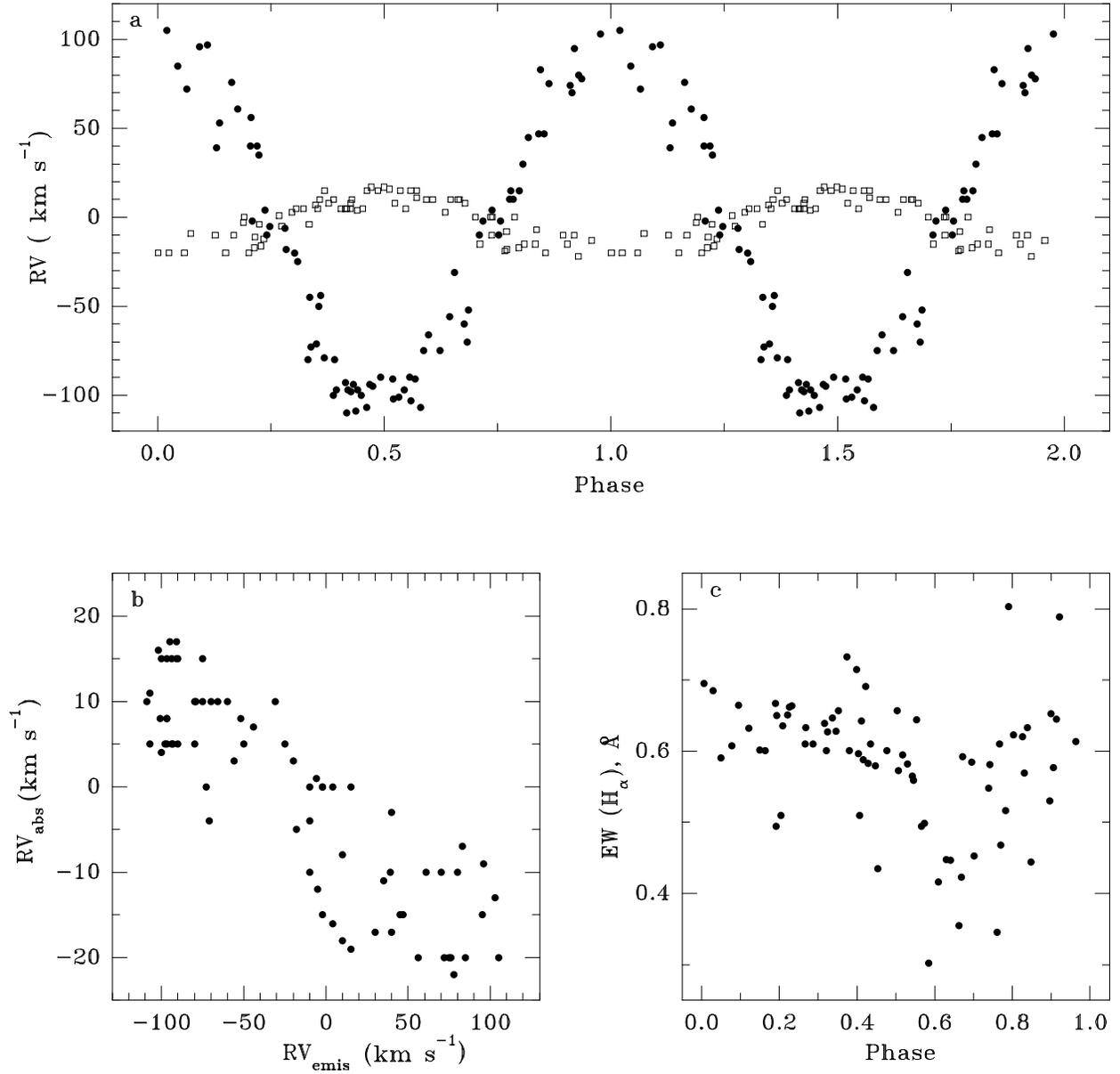}
\caption{The H$\alpha$ line radial velocity and
equivalent width phase curve.  a) The RVs of the traveling emission
component are shown by filled circles, while the RVs of the absorption
component are shown by open squares.  The mean error of the
measurements is 5~km~s$^{-1}$.
b) The RVs of the H$\alpha$ components plotted against each other.
c) The equivalent width of the traveling emission component.  \label{f7}}
\end{figure}

\clearpage

\begin{figure}
\epsscale{0.5}
\plotone{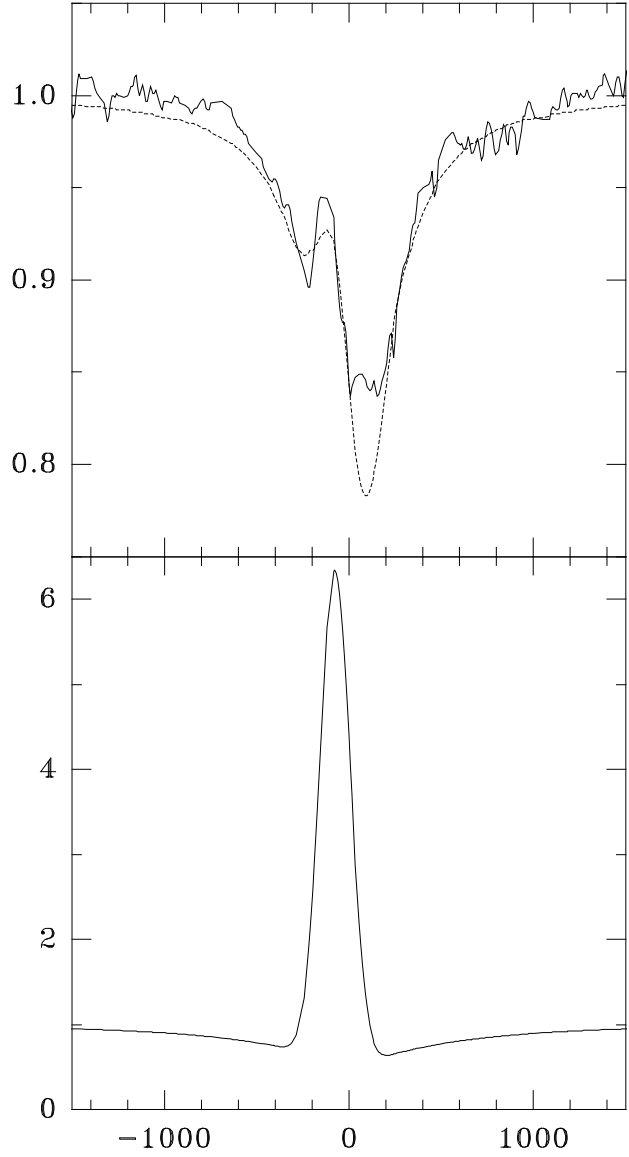}
\caption{The H$\alpha$ line profile obtained on 1996
November 4 and its theoretical fit.  The observed profile (solid line)
and the composite calculated profile (dashed line) are shown in the
upper panel. The composite profile consists of two \citet{k94}
theoretical profiles, ($T_{\rm eff}=25,000$~K, $\log~g=4.0$ and
$T_{\rm eff}=9,000$~K, $\log~g=4.0$), and a Gaussian emission. The
continuum brightness ratio is 40. Both photospheric profiles are
broadened at $v\,\sin\,i=250~{\rm km~s^{-1}}$. The lower panel represents
the appearance of the Gaussian emission in the secondary spectrum. Both
composite profiles are normalized to the underlying continuum. The
velocity scale is in km~s$^{-1}$. \label{f8}}
\end{figure}

\clearpage

%%%%%%%%%%%%%%% Table 1 %%%%%%%%%%%%%%%%%%%%%%
\renewcommand{\arraystretch}{1.00}
\begin{table}
\begin{center}
{\scriptsize
\caption{Summary of the spectroscopic observations of $\pi$~Aqr \label{t1}}
\begin{tabular}{ccrcrccrcr}
\tableline\tableline
JD       & EW    & V$_{\rm em}$ &I$_{\rm em}$ & V$_{\rm abs}$&
JD       & EW    & V$_{\rm em}$ &I$_{\rm em}$ & V$_{\rm abs}$\\
2400000+ &  \AA  & km~s$^{-1}$  &             & km~s$^{-1}$  &
2400000+ &  \AA  & km~s$^{-1}$  &             & km~s$^{-1}$  \\
(1) & (2) & (3) & (4) & (5) & (1) & (2) & (3) & (4) & (5) \\
\tableline
50351.73 & 0.65 &    70& 0.13 &$-$10 & 51480.59 & 0.63 &$-$80 & 0.10 &5    \\
50357.08 & 0.60 &   103& 0.12 &$-$13 & 51483.57 & 0.64 &$-$79 & 0.10 &10   \\
50360.67 & 0.66 &   105& 0.12 &$-$20 & 51487.54 & 0.69 &$-$93 & 0.10 &5    \\
50362.66 & 0.68 &    85& 0.12 &$-$20 & 51488.55 & 0.62 &$-$98 & 0.10 &5    \\
50366.70 & 0.62 &    96& 0.11 &$-$9  & 51489.51 & 0.65 &$-$109& 0.10 &10   \\
50391.56 & 0.68 &$-$100& 0.12 &15    & 51490.52 & 0.61 &$-$100& 0.09 &4    \\
50402.53 & 0.64 &$-$91 & 0.09 &17    & 51491.54 & 0.57 &$-$107& 0.09 &5    \\
50590.86 & 0.54 &$-$2  & 0.09 &0     & 51496.48 & 0.55 &$-$102& 0.08 &16   \\
50628.87 & 0.58 &$-$2  & 0.11 &0     & 51497.50 & 0.56 &$-$101& 0.09 &8    \\
50639.83 & 0.57 &$-$73 & 0.09 &0     & 51498.51 & 0.54 &$-$97 & 0.09 &15   \\
50640.84 & 0.65 &$-$71 & 0.11 &$-$4  & 51499.51 & 0.51 &$-$90 & 0.07 &5    \\
50660.79 & 0.46 &$-$75 & 0.07 &15    & 51500.50 & 0.57 &$-$91 & 0.08 &15   \\
50663.82 & 0.40 &$-$75 & 0.08 &10    & 51501.53 & 0.45 &$-$107& 0.08 &11   \\
50668.83 & 0.42 &$-$70 & 0.06 &10    & 51510.46 & 0.55 &$-$52 & 0.07 &8    \\
50674.74 & 0.54 &$-$10 & 0.06 &$-$10 & 51512.45 & 0.55 &$-$10 & 0.07 &0    \\
50688.78 & 0.56 &   95 & 0.08 &$-$15 & 51520.51 & 0.78 &   30 & 0.08 &$-$17\\
50704.71 & 0.63 &   97 & 0.08 &$-$   & 51521.52 & 0.62 &   45 & 0.07 &$-$15\\
50712.73 & 0.62 &   40 & 0.08 &$-$3  & 51523.51 & 0.60 &   47 & 0.08 &$-$15\\
50715.68 & 0.60 &$-$10 & 0.07 &$-$4  & 51524.50 & 0.60 &   47 & 0.08 &$-$10\\
50728.66 & 0.60 &$-$97 & 0.10 &8     & 51531.51 & 0.77 &   78 & 0.11 &$-$22\\
50730.61 & 0.60 &$-$110& 0.08 &$-$   & 51777.71 & 0.44 &   75 & 0.06 &$-$20\\
50732.65 & 0.56 &$-$97 & 0.09 &8     & 51781.71 & 0.53 &   74 & 0.08 &$-$  \\
50742.59 & 0.54 &$-$103& 0.08 &$-$   & 51794.72 & 0.59 &   72 & 0.09 &$-$20\\
50761.57 & 0.44 &   10 & 0.05 &$-$8  & 51800.69 & 0.63 &   53 & 0.10 &$-$  \\
51004.83 & 0.34 &$-$60 & 0.07 &10    & 51806.64 & 0.49 &   56 & 0.09 &$-$20\\
51045.79 & 0.57 &   76 & 0.09 &$-$20 & 51807.69 & 0.51 &   40 & 0.08 &$-$17\\
51052.84 & 0.64 &$-$5  & 0.09 &$-$12 & 51814.68 & 0.61 &$-$20 & 0.09 &3    \\
51055.72 & 0.60 &$-$6  & 0.08 &1     & 51824.62 & 0.51 &$-$97 & 0.09 &5    \\
51097.63 & 0.59 &   15 & 0.07 &$-$19 & 51825.56 & 0.59 &$-$94 & 0.14 &5    \\
51146.50 & 0.63 &$-$44 & 0.10 &7     & 51828.64 & 0.44 &$-$94 & 0.10 &15   \\
51396.79 & 0.61 &$-$45 & 0.09 &$-$   & 51830.64 & 0.60 &$-$90 & 0.10 &15   \\
51422.74 & 0.43 &$-$56 & 0.07 &3     & 51839.61 & 0.30 &$-$66 & 0.08 &10   \\
51423.70 & 0.43 &$-$31 & 0.06 &10    & 51849.59 & 0.45 &$-$2  & 0.10 &$-$15\\
51435.71 & 0.48 &   15 & 0.07 &0     & 51854.56 & 0.35 &   10 & 0.06 &$-$18\\
51439.70 & 0.53 &   83 & 0.06 &$-$7  & 52103.81 & 0.51 &    4 & 0.08 &0    \\
51446.66 & 0.61 &   80 & 0.11 &$-$10 & 52136.72 & 0.68 &   39 & 0.10 &$-$10\\
51467.64 & 0.57 &   61 & 0.10 &$-$10 & 52151.73 & 0.62 &$-$25 & 0.10 &5    \\
51471.56 & 0.63 &   35 & 0.10 &$-$11 & 52155.77 & 0.68 &$-$50 & 0.09 &5    \\
51472.64 & 0.62 &    4 & 0.09 &$-$16 & 52158.65 & 0.60 &$-$80 & 0.10 &10   \\
51476.57 & 0.61 &$-$18 & 0.08 &$-$5  & 52165.67 & 0.66 &$-$95 & 0.10 &17   \\
\tableline
\end{tabular}
\vspace{-11.0mm}
\tablecomments{The heliocentric Julian date for the observation is listed in
column~1; EW (column~2) is the equivalent width of the emission component of
the H$\alpha$ profile (a theoretical profile for $T_{\rm eff}=24,000$~K,
$\log~g=4.0$, $v\,sin\,i=250~{\rm km~s^{-1}}$ is subtracted);
$V_{\rm em}$ (column~3) is the heliocentric velocity of the traveling
emission peak and $I_{\rm em}$ (column~4) is its the residual intensity in
continuum units after subtraction of a theoretical photospheric profile;
$V_{\rm abs}$ (column~5) is the heliocentric velocity of the absorption
part of the H$\alpha$ profile.}
}
\end{center}
\end{table}
%%%%%%%%%%%%%%%%%%%%%%%%%%%%%%%%%%%%%%%%%%%%%%%%%%%%%%%%%%%%%%

\clearpage

%%%%%%%%%%%%%%% Table 2 %%%%%%%%%%%%%%%%%%%%%%%%%%%%%%%%%%%%%%
\begin{table}
\begin{center}
\caption[ ] {Parameters of the RV curves for the H$\alpha$ line components \label{t2}}
\begin{tabular}{cclccclr}
\tableline
\tableline
\noalign{\smallskip}
Component & $\gamma$    &   $K$        & $P$  &    $t_0$   &r.m.s       \\
          & km~s$^{-1}$ & km~s$^{-1}$  & days & JD2450000+ &km~s$^{-1}$ \\
\noalign{\smallskip}
\tableline
\noalign{\smallskip}
emission  & $-1.6\pm$0.1& 101.4$\pm$0.2 & 84.135$\pm$0.004 & 274.84$\pm$0.04 & 14.4\\
absorption& $-4.9\pm$0.1&  16.7$\pm$0.2 & 84.07$\pm$0.02   & 276.5$\pm$13.2  &  6.0\\
\noalign{\smallskip}
\tableline
\end{tabular}
\end{center}
\end{table}
%%%%%%%%%%%%%%%%%%%%%%%%%%%%%%%%%%%%%%%%%%%%%%%%%%%%%%%%%%%%%%

\end{document}